\documentclass[amstex,preprint,eqsecnum,aps]{revtex4}

\begin{document}

\title{Is there Unruh radiation?}

\author{G. W. Ford}

\affiliation{Department of Physics, University of
Michigan, \\ Ann Arbor, Michigan 48109-1120}

\author{R. F. O'Connell}

\affiliation{Department of Physics and Astronomy,
Louisiana State University, \\ Baton Rouge, Louisiana 70803-4001}

\date{\today}

\begin{abstract}
It is generally accepted that a system undergoing uniform acceleration with
respect to zero-temperature vacuum will thermalize at a finite temperature
(the so-called Unruh temperature) that is proportional to the acceleration.
However, the question of whether or not the system actually radiates is
highly controversial. Thus, we are motivated to present an exact calculation
using a generalized quantum Langevin equation to describe an oscillator (the
detector) moving under a constant force and coupled to a one-dimensional
scalar field (scalar electrodynamics). Moreover, our analysis is simplified by 
using the oscillator as a detector. We show that this system does not radiate
despite the fact that it does in fact thermalize at the Unruh temperature.  We
remark upon a differing opinion expressed regarding a system coupled to the
electromagnetic field.
\end{abstract}


\maketitle

\section{Introduction}
\label{sec:one}

It is now generally accepted that, as originally pointed out by Davies \cite
{davies75} and Unruh \cite{unruh76}, a system undergoing uniform
acceleration with respect to zero-temperature vacuum will come to
equilibrium at an effective temperature that is proportional to the
acceleration. This is the so-called Unruh temperature. What is more
controversial is whether or not this implies that the system actually
radiates. 

Grove \cite{grove86} was the first to argue that, contrary to the then
prevailing opinion, the system does not radiate. This conclusion was
supported by Raine et al. \cite{raine91} who considered a uniformly
accelerated oscillator moving under the action of a constant force and
analyzed its effect on a detector (represented as an inertial harmonic
oscillator). However, Unruh \cite{unruh92} claims that Raine et al. \cite
{raine91} discarded some terms in the autocorrelation function of the field
that actually contribute to the excitation of the detector. Also, we note
that Ref. \cite{raine91} uses a Weisskopf-Wigner (or white- noise)
approximation. Belief in the reality of the radiation may be gauged by
recent suggestions as to how it might be measured \cite{chen99, scully03},
but there is widespread controversy \cite{hu04} as to whether the radiation
actually exists. Thus, we are motivated to present an exact calculation for
the simple model of an oscillator (the detector) coupled to a scalar field
(scalar electrodynamics). In particular we discard no terms and do not
introduce the Weisskopf-Wigner approximation. Our approach is also unique in
using the oscillator as a detector since it considerably simplifies the
analysis in that we only need to treat the motion of one body instead of two. 
Secondly, making the oscillator massive enough has the merit of ensuring that
the back reaction due to the scalar field has no effect on the oscillators
dynamics. The methods we use are those of a quantum Langevin approach which we
have used for such problems as an oscillator coupled to the radiation field
\cite{ford88b}.

Since the subject has given rise to so much controversy, our aim will be to
present the discussion in a detailed pedagogical manner. We begin in Sec. 
\ref{sec:two}, where we give a simple description of the real scalar field
in one dimension. This field is isomorphic to the case of a stretched string 
\cite{ford88b} and to make the discussion more intuitive we couch it in
terms of the string. Our starting- point is the Lagrangian for the field,
from which we deduce that the equation of motion of the field and the zero-
temperature correlation are invariant under Lorentz transformation. In
addition, we obtain explicit expressions for the correlation (which consists
of a finite space and time dependent part plus a constant divergent term)
and the commutator

In Sec. \ref{sec:three}, we consider the case of a point mass moving under a
constant force (hyperbolic motion). There we make use of the convenient
parametric representation of the motion sometimes called Rindler coordinates 
\cite{rindler91}. We find that the zero-temperature
correlation of the field evaluated at a space-time point moving along the
hyperbolic path is identical with the correlation evaluated at a point fixed
in space, but with the field at the elevated Unruh temperature.

Next, we take into account the coupling of an oscillator to the field. In
Sec. \ref{sec:four}, we consider the coupling of a charged oscillator to the
field (scalar electrodynamics), such that the oscillator is at rest. This
system is similar to the Lamb model of a particle attached to a stretched
string \cite{ford88b}, in that it leads to the same Langevin equation. The
solution of the Langevin equation enables us to calculate the field and with
that the flux of energy radiated by the oscillator at rest and in
equilibrium with the field at temperature $T$. We find the net energy flux
at any point in the field is identically zero, the result of a detailed
balance of a flux of field energy emitted by the oscillator and a flux of
field energy supplied to the oscillator. This is entirely what one should
expect, since the mean energy of the oscillator in equilibrium is constant.
The point of this exercise is seen in the next two sections, where we find
the for an oscillator in hyperbolic motion through a zero temperature field
the net flux vanishes in the same way and for the same reasons as for an
oscillator at rest.

In Sec. \ref{sec:five}, we extend the model discussed in the previous
Section to consider an oscillator coupled to a moving point in the field. In
contrast to Raine et al. who introduced an inertial detector at a fixed
point in space to test for the emission of radiation from the moving
oscillator, we simply treat the oscillator as a detector and calculate the
flux of field energy. In addition, we do not ignore quantum effects in the
expressions for the various field correlation functions. As before, we
obtain the Unruh temperature. Then in Sec. \ref{sec:six} we present an
explicit calculation to show that for an oscillator in hyperbolic motion the
expectation value of the energy flux vanishes, just as for an oscillator at
rest. With some concluding remarks in Sec. \ref{sec:seven}, we present our
conclusion that, whereas one can speak of an Unruh temperature, there is no
corresponding radiation to be detected.  In this context, we also analyze Unruh's
counterclaim [5] and argue that it is not valid.  Finally, we emphasize that our
discussion is restricted to the specific model of an oscillator coupled to a
one-dimensional scalar field.  While this is the model used by most authors,
including the original work of Davis \cite{davies75} and Unruh \cite{unruh76},
other models (for example the more realistic one of a charged particle coupled to
the electromagnetic field \cite{boyer84}) can give different results.  In our
concluding remarks we discuss conflicting opinions concerning the radiation with
such models.

\section{Real scalar field in one dimension (stretched string).}

\label{sec:two}

The Lagrangian for the stretched string is

\begin{equation}
L=\int dy{\LARGE \{}{\frac{\sigma }{2}}({\frac{\partial u}{\partial t}})^{2}-
{\frac{\tau }{2}}({\frac{\partial u}{\partial y}})^{2}{\LARGE \}},
\label{2.1}
\end{equation}

where $\sigma $ is the mass per unit length, $\tau $ is the tension and $
u(y,t)$ is the string displacement. The integral is along the length of the
string, which is stretched in the $y$ direction.. For the real scalar field
it is customary to put $\sigma =1/4\pi $ and $\tau =c^{2}/4\pi $, where $c$
is the velocity of light. In that case $u$ has the dimensions $(mass\cdot
length)^{1/2}$. The equation of motion of the string is the homogeneous wave
equation,

\begin{equation}
{\frac{\partial ^{2}u}{\partial t^{2}}}-c^{2}{\frac{\partial ^{2}u}{\partial
y^{2}}}=0,  \label{2.2}
\end{equation}

where $c$ is the speed of waves in the string, 

\begin{equation}
c=(\tau /\sigma )^{1/2}.  \label{2.3}
\end{equation}

There is an energy conservation law,

\begin{equation}
\frac{\partial \mathcal{E}}{\partial t}+\frac{\partial j}{\partial y}=0,
\label{2.4}
\end{equation}

with energy density

\begin{equation}
\mathcal{E}(y,t)=\frac{\sigma }{2}\left( \frac{\partial u(y,t)}{\partial t}
\right) ^{2}+\frac{\tau }{2}\left( {\frac{\partial u(y,t)}{\partial y}}
\right) ^{2}  \label{2.5}
\end{equation}

and energy flux

\begin{equation}
j(y,t)=-\frac{\tau }{2}\left( \frac{\partial u(y,t)}{\partial t}\frac{
\partial u(y,t)}{\partial y}+\frac{\partial u(y,t)}{\partial y}\frac{
\partial u(y,t)}{\partial t}\right) .  \label{2.6}
\end{equation}

Although for the string $c$ is not necessarily the speed of light, the
equation of motion is still invariant under the Lorentz transformation with
velocity $v$:

\begin{equation}
y^{\prime }=\gamma (y-vt),\qquad t^{\prime }=\gamma (t-vy/c^{2}),\qquad
u^{\prime }(y^{\prime },t^{\prime })=u(y,t),  \label{2.7}
\end{equation}

where 

\begin{equation}
\gamma =(1-v^{2}/c^{2})^{-1/2}.  \label{2.8}
\end{equation}

Of course, for the real scalar field in one dimension this is the usual
Lorentz transformation.

The normal mode expansion of the displacement operator may be written
 
\begin{equation}
u(y,t)=\sum_{k}\sqrt{{\frac{\hbar }{2\sigma L\omega }}}(a_{k}e^{i(ky-\omega
t)}+a_{k}^{\dag }e^{-i(ky-\omega t)}),  \label{2.9}
\end{equation}

where $L$ is the length of the string and for periodic boundary conditions
the sum is over positive and negative integer multiples of $2\pi /L$. The
frequency is given by the dispersion relation,

\begin{equation}
\omega =c\left\vert k\right\vert .  \label{2.10}
\end{equation}

The string is quantized when we assume the canonical commutation relations
for the dimensionless normal mode amplitudes,

\begin{equation}
\lbrack a_{k},a_{k^{\prime }}^{\dag }]=\delta _{k^{\prime },k},\qquad
\lbrack a_{k},a_{k^{\prime }}]=0.  \label{2.11}
\end{equation}

When the string is in equilibrium at temperature $T$, we have the
expectation values

\begin{eqnarray}
\left\langle a_{k}a_{k^{\prime }}^{\dag }+a_{k^{\prime }}^{\dag
}a_{k}\right\rangle &=&\coth \frac{\hbar \omega }{2kT}\delta _{k^{\prime
},k},  \nonumber \\
\left\langle a_{k}a_{k^{\prime }}+a_{k^{\prime }}a_{k}\right\rangle &=&0.
\label{2.12}
\end{eqnarray}

The correlation function for the string is

\begin{equation}
C(\Delta y,\Delta t)\equiv {\frac{1}{2}}\left\langle
u(y_{1},t_{1})u(y_{2},t_{2})+u(y_{2},t_{2})u(y_{1},t_{1})\right\rangle
\label{2.13}
\end{equation}

where 

\begin{equation}
\Delta y=y_{1}-y_{2},\qquad \Delta t=t_{1}-t_{2}.  \label{2.14}
\end{equation}

In the limit of an infinite string ($L\rightarrow \infty $) we evaluate this
using the above relations together with the prescription

\begin{equation}
\sum_{k}\rightarrow \frac{L}{2\pi }\int_{-\infty }^{\infty }dk.  \label{2.15}
\end{equation}

The result is

\begin{equation}
C(\Delta y,\Delta t)=\frac{\hbar }{4\pi \sigma }\int_{-\infty }^{\infty }dk
\frac{1}{\omega }\coth \frac{\hbar \omega }{2kT}\cos \left( k\Delta y-\omega
\Delta t\right) .  \label{2.16}
\end{equation}

This integral is divergent at long wavelength ($k=0$). This is to be
expected since the Lagrangian (\ref{2.1}) is invariant under uniform
displacement of the string. Nevertheless, we can still obtain a useful
result. To do so note that the derivative is a conditionally convergent
integral \cite{bateman_it1},

\begin{eqnarray}
\frac{\partial C(\Delta y,\Delta t)}{\partial \Delta t} &=&-\frac{\hbar }{
4\pi \sigma c}\int_{0}^{\infty }d\omega \coth \frac{\hbar \omega }{2kT}[\sin
\omega (\Delta t-\frac{\Delta y}{c})+\sin \omega (\Delta t+\frac{\Delta y}{c}
)]  \nonumber \\
&=&-\frac{kT}{4\sigma c}(\coth \frac{\pi kT(\Delta t-\Delta y/c)}{\hbar}
+\coth \frac{\pi kT(\Delta t+\Delta y/c)}{\hbar}).  \label{2.17}
\end{eqnarray}

From this we conclude that

\begin{eqnarray}
C(\Delta y,\Delta t) &=&-\frac{\hbar }{4\pi \sigma c}(\log \sinh \frac{\pi
kT(\Delta t-\Delta y/c)}{\hbar }+\log \sinh \frac{\pi kT(\Delta t+\Delta y/c)
}{\hbar }\}  \nonumber \\
&&+\text{constant},  \label{2.18}
\end{eqnarray}

where the constant, while infinite, is independent of $\Delta y$ and $\Delta
t$. The case of the correlation at a fixed \ point on the string ($\Delta
y=0 $) is of special interest,

\begin{equation}
C(0,\Delta t)=-\frac{\hbar }{2\pi \sigma c}\log \sinh \frac{\pi kT\Delta t}{
\hbar }+\text{constant.}  \label{2.19}
\end{equation}

This finite temperature correlation function in Minkowski space-time will in
the next section be compared to the zero-temperature correlation function in
Rindler coordinates in order to relate the constant acceleration to the
Unruh temperature.

It is of interest to consider the zero-temperature correlation function in
Minkowski space, 

\begin{equation}
C_{0}(\Delta y,\Delta t)\equiv {\frac{\hbar }{4\pi \sigma }}\int_{-\infty
}^{\infty }\frac{dk}{\omega }\cos (k\Delta y-\omega \Delta t).  \label{2.20}
\end{equation}

Introduce in the integral a Lorentz transformation of the wave vector and
frequency, 

\begin{equation}
k^{\prime }=\gamma (k-v\omega /c^{2}),\qquad \omega ^{\prime }=\gamma
(\omega -vk).  \label{2.21}
\end{equation}

It is a simple matter to show that the dispersion relation (\ref{2.10}) is
preserved under this transformation and that

\begin{equation}
dk^{\prime }/\omega ^{\prime }=dk/\omega .  \label{2.22}
\end{equation}

To obtain an explicit expression for the zero-temperature correlation, put $
v=-\Delta y/\Delta t$ if $\left\vert \Delta y/\Delta t\right\vert <c$ and $\
v=-c^{2}\Delta t/\Delta y$ if $\left\vert \Delta y/\Delta t\right\vert >c$ \
Then, using the dispersion relation (\ref{2.10}), we obtain the expression
 
\begin{equation}
C_{0}(\Delta y,\Delta t)={\frac{\hbar }{2\pi \sigma c}}\int_{0}^{\infty
}d\omega {\frac{\cos (\omega \left\vert \Delta t^{2}-\Delta
y^{2}/c^{2}\right\vert ^{1/2})}{\omega }}.  \label{2.23}
\end{equation}

This integral is again divergent at long wavelength. Note that the
divergence comes from the behavior at $\omega =0$ and that we can write 

\begin{equation}
C_{0}(\Delta y,\Delta t)=\lim_{\epsilon \rightarrow 0^{+}}{\frac{\hbar }{
2\pi \sigma c}}\int_{\epsilon (\Delta t^{2}-\Delta
y^{2}/c^{2})^{1/2}}^{\infty }dt{\frac{\cos (t)}{t}}.  \label{2.24}
\end{equation}

Here, the integral is logarithmic at $t=0$, so we see that
 
\begin{equation}
C_{0}(\Delta y,\Delta t)=-{\frac{\hbar }{4\pi \sigma c}}\log \left\vert
\Delta t^{2}-\frac{\Delta y^{2}}{c^{2}}\right\vert +\text{constant},
\label{2.25}
\end{equation}

where the constant is logarithmically divergent as $\epsilon \rightarrow 0$,
but independent of $\Delta t$ and $\Delta y$. Note that, since it depends
only upon the invariant interval, this zero-temperature correlation is
invariant under Lorentz transformation

Next consider the commutator of the field, which can be written 

\begin{equation}
\lbrack u(y_{1},t_{1}),u(y_{2},t_{2})]={\frac{i\hbar }{2\pi \sigma }}
\int_{-\infty }^{\infty }dk{\frac{\sin (k\Delta y-\omega \Delta t)}{\omega }}. 
\label{2.26}
\end{equation}

If $\left\vert \Delta y/\Delta t\right\vert <c$, we introduce a Lorentz
transformation corresponding to $v=-\Delta y/\Delta t$ to obtain the
expression 

\begin{equation}
\lbrack u(y_{1},t_{1}),u(y_{2},t_{2})]={\frac{-i\hbar }{2\pi \sigma }}
\int_{-\infty }^{\infty }dk{\frac{\sin [\omega \Delta t(1-\Delta
y^{2}/c^{2}\Delta t^{2})^{1/2}]}{\omega }}={\frac{-i\hbar }{2\sigma c}}\text{
sgn}(\Delta t).  \label{2.27}
\end{equation}

On the other hand, if $\left\vert \Delta y/\Delta t\right\vert >c$, we
introduce a Lorentz transformation corresponding to $v=-c^{2}\Delta t/\Delta
y$ to obtain the expression 

\begin{equation}
\lbrack u(y_{1},t_{1}),u(y_{2},t_{2})]={\frac{-i\hbar }{2\pi \sigma }}
\int_{-\infty }^{\infty }dk{\frac{\sin [k\Delta y(1-c^{2}\Delta t^{2}/\Delta
y^{2})^{1/2}]}{\omega }}=0.  \label{2.28}
\end{equation}

Therefore, we have the general result 

\begin{equation}
\lbrack u(y_{1},t_{1}),u(y_{2},t_{2})]={\frac{\hbar }{2i\sigma c}}\text{sgn}
(\Delta t)\theta (\Delta t^{2}-\frac{\Delta y^{2}}{c^{2}}),  \label{2.29}
\end{equation}

in which $\theta $ is the Heaviside function. Note that the commutator is
invariant under proper Lorentz transformation, with an extra sign change
under time reversal.

\section{Hyperbolic motion.}

\label{sec:three}

A point mass $m$ moving under a constant force $F$ in relativity moves
according to the equation of motion 

\begin{equation}
{\frac{d}{dt}}{\frac{mv}{(1-v^{2}/c^{2})^{1/2}}}=F,  \label{3.1}
\end{equation}

where $v=dy/dt$ is the velocity. The solution can most simply be written in
parametric form,

\begin{equation}
y={\frac{mc^{2}}{F}}\cosh \frac{F\tau }{mc},\quad t={\frac{mc}{F}}\sinh 
\frac{F\tau }{mc}\qquad -\infty <\tau <\infty ,  \label{3.2}
\end{equation}

where the parameter $\tau $ is the proper time, 

\begin{equation}
d\tau =(1-v^{2}/c^{2})^{1/2}dt.  \label{3.3}
\end{equation}

This solution is called hyperbolic motion. It is also called uniformly
accelerated motion, since in the instantaneous rest frame the acceleration
is a constant equal to $F/m.$The motion corresponds to a point mass coming
at $t=-\infty $ from $y=+\infty $ with velocity $v=-c$, decelerating with a
constant force $F$ until at $t=0$ it comes to rest at $y=mc^{2}/F$. The mass
then accelerates back to $y=+\infty $ at $t=\infty $.

For this hyperbolic motion, taking the two points to be on the same world
line (\ref{3.2}), we see that 

\begin{equation}
(\Delta t^{2}-\Delta y^{2}/c^{2})^{1/2}={\frac{2mc}{F}}\sinh \frac{F\Delta
\tau }{2mc},  \label{3.4}
\end{equation}

where $\Delta \tau =\tau _{1}-\tau _{2}$. The zero-temperature correlation
function (\ref{2.25}) for the scalar field in one dimension, when evaluated
on the world line, therefore takes the form

\begin{equation}
C_{0}(\Delta y,\Delta t)=-{\frac{\hbar }{2\pi \sigma c}}\log \sinh \frac{%
F\Delta \tau }{2mc}+\text{constant.}  \label{3.5}
\end{equation}

Therefore, for hyperbolic motion the correlation is a function of $\Delta
\tau $ alone.

Recall that the proper time $\tau $ is the time as measured on a moving
clock. Therefore, comparing (\ref{3.5}) and (\ref{2.19}), we see that the
the zero-temperature correlation evaluated along the hyperbolic path is
identical to the finite temperature correlation evaluated at a fixed point
if we make the identification 

\begin{equation}
kT={\frac{\hbar F}{2\pi mc}}.  \label{3.6}
\end{equation}

This is the Unruh temperature \cite{unruh76}.

\section{Oscillator coupled to a one-dimensional scalar field.}

\label{sec:four}

We consider a coupling of the oscillator to the field through the velocity.
The Lagrangian is 

\begin{equation}
\mathit{L}={\frac{1}{2}}mv^{2}-{\frac{1}{2}}Kx^{2}-2\sigma cv\phi (0,t)+\int
dy\left\{ {\frac{\sigma }{2}}(\frac{\partial \phi }{\partial t})^{2}-{\frac{
\sigma c^{2}}{2}}({\frac{\partial \phi }{\partial y}})^{2}\right\} .
\label{4.1}
\end{equation}

This is sometimes called scalar electrodynamics. Note that the particle
displacement is in the $x$ direction, while the field extends in the $y$
direction. The oscillator interacts with the field at the origin ($y=0$).
Thus, the system is very like the Lamb model (in which the particle is
attached to the center of an infinite stretched string \cite{lamb1900,ford88b})
and we shall see that it leads to the same quantum Langevin equation. However, if
the system is to be invariant under time reversal then $\phi $ must be odd under
time reversal. In this sense, the field $\phi $ is different from the
displacement $u$ of a string. Otherwise the discussion of the previous sections
applies to the free field $\phi $.

The equation of particle motion is that of a driven oscillator,

\begin{equation}
m{\frac{d^{2}x}{dt^{2}}}+Kx=2\sigma c{\frac{\partial \phi (0,t)}{\partial t}}.  \label{4.2}
\end{equation}

The field equation of motion is the inhomogeneous wave equation,

\begin{equation}
{\frac{\partial ^{2}\phi }{\partial t^{2}}}-c^{2}{\frac{\partial ^{2}\phi }{
\partial y^{2}}}=-2c{\frac{dx(t)}{dt}}\delta (y).  \label{4.3}
\end{equation}

We now eliminate the field variable between these two equations. Treating
the right hand side as known, the solution of the field equation is 

\begin{equation}
\phi (y,t)=\phi ^{h}(y,t)-x(t-\frac{\left\vert y\right\vert }{c}),
\label{4.4}
\end{equation}

where $\phi ^{h}(y,t)$ is the general solution of the homogeneous wave
equation (\ref{2.2}). Putting this solution in the particle equation of
motion, we get the Langevin equation 

\begin{equation}
m\ddot{x}+\zeta {\dot{x}}+Kx=F(t),  \label{4.5}
\end{equation}

where 

\begin{equation}
\zeta =2\sqrt{\sigma \tau }  \label{4.6}
\end{equation}

is the friction constant and 

\begin{equation}
F(t)=\zeta {\frac{\partial \phi ^{h}(0,t)}{\partial t}}  \label{4.7}
\end{equation}

is a fluctuating force operator.

The free field has the normal mode expansion (\ref{2.9}),

\begin{equation}
\phi ^{h}(y,t)=\sum_{k}\sqrt{{\frac{\hbar c}{\zeta L\omega }}}
(a_{k}e^{i(ky-\omega t)}+a_{k}^{\dag }e^{-i(ky-\omega t)}),  \label{4.8}
\end{equation}

in which we combined (\ref{2.3}) and (\ref{4.6}) to write $\zeta =2\sigma c$
. From this we see that the fluctuating force can be expanded, 

\begin{equation}
F(t)=\sum_{k}\sqrt{{\frac{\hbar c\zeta \omega }{L}}}(-ia_{k}e^{-i\omega
t}+ia_{k}^{\dag }e^{i\omega t}).  \label{4.9}
\end{equation}

From this, using the canonical commutation rules (\ref{2.11}) and the
expectation values (\ref{2.12}), we can obtain the correlation and
commutator for the fluctuating force. If we then form the limit $
L\rightarrow \infty $, using the prescription (\ref{2.15}) and the
dispersion relation (\ref{2.10}), we get

\begin{eqnarray}
{\frac{1}{2}}\left\langle F(t_{1})F(t_{2})+F(t_{2})F(t_{1})\right\rangle &=&{
\frac{\hbar }{\pi }}\int_{0}^{\infty }d\omega \zeta \omega \coth {\frac{
\hbar \omega }{2kT}}\cos [\omega (t_{1}-t_{2})],  \nonumber \\
\lbrack F(t_{1}),F(t_{2})] &=&-i{\frac{2\hbar }{\pi }}\int_{0}^{\infty
}d\omega \zeta \omega \sin [\omega (t_{1}-t_{2})].  \label{4.10}
\end{eqnarray}

Using the expansion (\ref{4.9}) of the fluctuating force in the right hand
side of the Langevin equation (\ref{4.5}), we see that the solution has the
expansion

\begin{equation}
x(t)=\sum_{k}\sqrt{\frac{\hbar c\zeta \omega }{L}}[-i\alpha (\omega
)a_{k}e^{-i\omega t}+i\alpha (\omega )^{\ast }a_{k}^{\dag }e^{i\omega t}].
\label{4.11}
\end{equation}

where $\alpha (\omega )$ is the oscillator susceptibility,

\begin{equation}
\alpha (\omega )=(-m\omega ^{2}-i\omega \zeta +K)^{-1}.  \label{4.12}
\end{equation}

It is of interest to form the position correlation for the oscillator. Using
the expectation values (\ref{2.12}) and forming the limit $L\rightarrow
\infty $, we find

\begin{equation}
{\frac{1}{2}}\left\langle x(t_{1})x(t_{2})+x(t_{2})x(t_{1})\right\rangle =
\frac{\hbar }{\pi }\int_{0}^{\infty }d\omega \mathrm{Im}\{\alpha (\omega
)\}\coth \frac{\hbar \omega }{2kT}\cos [\omega (t_{1}-t_{2})],  \label{4.13}
\end{equation}

in which we have used the fact that $\mathrm{Im}\{\alpha (\omega )\}=\zeta
\omega \left\vert \alpha (\omega )\right\vert ^{2}$\

Consider now the flux of field energy radiated by the oscillator as measured
at some point $y$ away from the origin. The energy flux operator is given by
(\ref{2.6}) with $u(y,t)\rightarrow \phi (y,t)$, the total field given by
the solution (\ref{4.4}) of the field equation. With this, forming the
expectation, we can write

\begin{equation}
\left\langle j(y,t)\right\rangle =\left\langle j_{0}(y,t)\right\rangle
+\left\langle j_{\mathrm{dir}}(y,t)\right\rangle +\left\langle j_{\mathrm{int
}}(y,t)\right\rangle ,  \label{4.14}
\end{equation}

where $\left\langle j_{0}(y,t)\right\rangle $ is the energy flux in the
absence of the oscillator,

\begin{equation}
\left\langle j_{0}(y,t)\right\rangle =-\frac{1}{2}\zeta c\mathrm{Re}
\{\left\langle \frac{\partial \phi ^{h}(y,t)}{\partial t}\frac{\partial \phi
^{h}(y,t)}{\partial y}\right\rangle \},  \label{4.15}
\end{equation}

$\left\langle j_{\mathrm{dir}}(y,t)\right\rangle $ is the energy flux
arising from source alone,

\begin{equation}
\left\langle j_{\mathrm{dir}}(y,t)\right\rangle =\frac{1}{2}\frac{y}{
\left\vert y\right\vert }\zeta \left\langle \dot{x}^{2}(t-\frac{\left\vert
y\right\vert }{c})\right\rangle ,  \label{4.16}
\end{equation}

and $\left\langle j_{\mathrm{int}}(y,t)\right\rangle $ is the interference
term

\begin{equation}
\left\langle j_{\mathrm{int}}(y,t)\right\rangle =\frac{1}{2}\zeta \mathrm{Re}
\{\left\langle \dot{x}(t-\frac{\left\vert y\right\vert }{c})\left( c\frac{
\partial \phi ^{h}(y,t)}{\partial y}-\frac{y}{\left\vert y\right\vert }\frac{
\partial \phi ^{h}(y,t)}{\partial t}\right) \right\rangle  \label{4.17}
\end{equation}

To get some insight into the significance of these terms, we multiply both
sides of the Langevin equation (\ref{4.5}) by $dx/dt$, symmetrize the
factors in each term and form the expectation of the resulting equation to
get the oscillator energy balance equation:

\begin{equation}
\frac{d}{dt}\left\langle \frac{1}{2}m\dot{x}^{2}+\frac{1}{2}
Kx^{2}\right\rangle +\zeta \left\langle \dot{x}^{2}\right\rangle =\frac{1}{2}
\left\langle \dot{x}F+F\dot{x}\right\rangle .  \label{4.18}
\end{equation}

Here the first term on the left is clearly the rate of change of the mean
oscillator energy. The second term is interpreted as the mean rate of
radiation of field energy by the oscillator, while the right hand side is
interpreted as the mean rate at which work is done on the oscillator by the
fluctuating force. Of course, in equilibrium the mean oscillator energy is
constant and the remaining two terms must balance. Now, the direct term (\ref
{4.16}) is clearly the energy flux corresponding to the radiated energy,
directed away from the oscillator with half to the left and half to the
right. Note, incidentally, that the radiated power $\zeta \left\langle \dot{x
}^{2}\right\rangle $ is the analog of the well known Larmor formula for the
power radiated by an oscillating electric dipole (proportional to the square
of the velocity rather than the square of the acceleration since the
coupling is to a scalar rather than a vector field).

We now evaluate these fluxes using the expansion (\ref{4.8}) for the free
field and the expansion (\ref{4.11}) for the oscillator displacement.
Consider first $\left\langle j_{0}(y,t)\right\rangle $, the current in the
absence of the oscillator. This, of course, must vanish on very general
grounds. In this case we see that when we insert the expansion for the free
field the result is a sum over $k$ of an odd function of $k$, which
vanishes. Next consider the direct flux (\ref{4.16}). Using the expansion (
\ref{4.11}) and the expectation values (\ref{2.12}), we find after a little
rearrangement

\begin{equation}
\left\langle j_{\mathrm{dir}}(y,t)\right\rangle =\frac{y}{\left\vert
y\right\vert }\frac{\hbar c}{2L}\sum_{k}\omega ^{3}\zeta ^{2}\left\vert
\alpha (\omega )\right\vert ^{2}\coth \frac{\hbar \omega }{2kT}.
\label{4.19}
\end{equation}

If we use the prescription (\ref{2.15}) to form the limit $L\rightarrow
\infty $, we can write

\begin{equation}
\left\langle j_{\mathrm{dir}}(y,t)\right\rangle =\frac{1}{2}\frac{y}{
\left\vert y\right\vert }\frac{\hbar }{\pi }\int_{0}^{\infty }d\omega \omega
^{3}\zeta ^{2}\left\vert \alpha (\omega )\right\vert ^{2}\coth \frac{\hbar
\omega }{2kT}.  \label{4.20}
\end{equation}

In the same way the interference term (\ref{4.17}) becomes

\begin{equation}
\left\langle j_{\mathrm{int}}(y,t)\right\rangle =\frac{y}{\left\vert
y\right\vert }\frac{\hbar c^{2}}{2L}\sum_{k}\mathrm{Re}\{i(\frac{ky}{
\left\vert y\right\vert }+\left\vert k\right\vert )\omega \zeta \alpha
(\omega )e^{-i(ky-\omega \left\vert y\right\vert /c)}\}\coth \frac{\hbar
\omega }{2kT}.  \label{4.21}
\end{equation}

Recalling that the sum is over positive and negative $k$, we discard terms
that are odd in $k$. The result, again after forming the limit $L\rightarrow
\infty $, can be written

\begin{equation}
\left\langle j_{\mathrm{int}}(y,t)\right\rangle =-\frac{1}{2}\frac{y}{
\left\vert y\right\vert }\frac{\hbar }{\pi }\int_{0}^{\infty }d\omega \omega
^{2}\zeta \mathrm{Im\{}\alpha (\omega )\}\coth \frac{\hbar \omega }{2kT}.
\label{4.22}
\end{equation}

But, as we see from (\ref{4.12}), $\mathrm{Im\{}\alpha (\omega )\}=\omega
\zeta \left\vert \alpha (\omega )\right\vert ^{2}$. Therefore, comparing the
expressions (\ref{4.20}) and (\ref{4.22}), we see that $\left\langle j_{
\mathrm{int}}(y,t)\right\rangle =-\left\langle j_{\mathrm{dir}
}(y,t)\right\rangle $ and

\begin{equation}
\left\langle j(y,t)\right\rangle =0.  \label{4.23}
\end{equation}

This result, which took some doing to obtain, should have been expected from
the beginning. After all, in equilibrium the mean energy of the oscillator
is constant, so the mean energy flux radiated into the field by the
oscillator must be balanced by a mean energy flux from the field into the
oscillator. This is just the result (\ref{4.23}). Put another way, we can
now interpret $\left\langle j_{\mathrm{int}}(y,t)\right\rangle $ as the
inward flux of field energy to balance the radiated power.

Finally, we remark on the situation when the oscillator is excited, say, by
an impulse applied at $t=0$. In this case there will be a mean motion
superposed on the random thermal motion of the oscillator. One can then
calculate the net radiated flux of energy using only the expression (\ref
{4.16}) for $\left\langle j_{\mathrm{dir}}(y,t)\right\rangle $, evaluated
for the mean motion. There will be no interference term since the mean
motion will be uncorrelated with the random motion of the field.

\section{Oscillator moving in the field.}

\label{sec:five}

Consider now an oscillator undergoing a given motion in the field direction 
\cite{raine91}. The idea is that in addition to the $x$-motion the
oscillator has a given $y$-motion,

\begin{equation}
y=y(\tau ),\qquad t=t(\tau ),  \label{5.1}
\end{equation}

where $\tau $ is the proper time,

\begin{equation}
d\tau =(dt^{2}-dy^{2}/c^{2})^{1/2}.  \label{5.2}
\end{equation}

We shall later take this to be the hyperbolic motion described in Section 
\ref{sec:three}, but for now we assume only that the motion is mechanically
allowed, so $\left\vert dy/dt\right\vert <c$. The Lagrangian (\ref{4.1})
must be modified to take the motion into account. First of all, consider the
kinetic energy, which must be replaced by the relativistic free particle
Lagrangian \cite{rindler91},

\begin{equation}
\mathrm{L}_{\mathrm{free}}=-mc^{2}\sqrt{1-\frac{1}{c^{2}}\left( \frac{dy}{dt}
\right) ^{2}-\frac{1}{c^{2}}\left( \frac{dx}{dt}\right) ^{2}}, \label{5.3}
\end{equation}

But the $x$-motion is nonrelativistic while the $y$-motion is arbitrary, so
we expand,

\begin{equation}
\mathrm{L}_{\mathrm{free}}\cong -mc^{2}\sqrt{1-\frac{1}{c^{2}}\left( \frac{dy
}{dt}\right) ^{2}}-\frac{1}{\sqrt{1-\frac{1}{c^{2}}\left( \frac{dy}{dt}
\right) ^{2}}}\frac{1}{2}m\left( \frac{dx}{dt}\right) ^{2}.  \label{5.4}
\end{equation}

We drop the first term, since the $y$-motion is given, and replace the
kinetic energy in the Lagrangian (\ref{4.1}) with the second term. Next
consider the potential energy, which must be multiplied by the time-dilation
factor $\sqrt{1-\frac{1}{c^{2}}\left( \frac{dy}{dt}\right) ^{2}}$ \ Finally,
the interaction must involve the field at the instantaneous position of the
particle. The resulting Lagrangian can be written%

\begin{eqnarray}
\mathrm{L} &=&\frac{dt}{d\tau }\frac{1}{2}m\left( \frac{dx}{dt}\right) ^{2}-
\frac{d\tau }{dt}{\frac{1}{2}}Kx^{2}-2\sigma c{\frac{dx}{dt}}\phi \lbrack
y(\tau ),t(\tau )]  \nonumber \\
&&+\int dy{\frac{\sigma }{2}}\left[ \left( {\frac{\partial \phi }{\partial t}
}\right) ^{2}-c^{2}\left( {\frac{\partial \phi }{\partial y}}\right) ^{2}
\right] ,  \label{5.5}
\end{eqnarray}

where we have used the definition (\ref{5.2}) of the proper time.

With this Lagrangian, the oscillator equation of motion is

\begin{equation}
m{\frac{d^{2}x}{d\tau ^{2}}}+Kx=2\sigma c{\frac{d\phi \lbrack y(\tau
),t(\tau )]}{d\tau }},  \label{5.6}
\end{equation}

while that for the field is the inhomogeneous wave equation,

\begin{equation}
{\frac{\partial ^{2}\phi }{\partial t^{2}}}-c^{2}{\frac{\partial ^{2}\phi }{
\partial y^{2}}}=-2c{\frac{dx}{dt}}\delta \lbrack y-y(\tau )].  \label{5.7}
\end{equation}

Treating the right hand side as known, the solution of this wave equation is 

\begin{equation}
\phi (y,t)=\phi ^{h}(y,t)-x(\tau ^{\mathrm{ret}}),  \label{5.8}
\end{equation}

where $\phi ^{h}(y,t)$ is the general solution (\ref{4.8}) of the free wave
equation and $\tau ^{\mathrm{ret}}$ is the retarded time. The retarded time
is defined implicitly as a function of the field point $(y,t)$ by the
relation 

\begin{equation}
t-t(\tau ^{\mathrm{ret}})=\left\vert y-y(\tau ^{\mathrm{ret}})\right\vert /c,
\label{5.9}
\end{equation}

and corresponds to the point on the mechanical path (\ref{5.1}) where it
pierces the backward light cone centered at the field point. Note in
particular that when the field point is on the mechanical path, then $\tau ^{
\mathrm{ret}}=\tau $. Thus, the solution (\ref{5.8}) of the inhomogeneous
wave equation can be written $\phi \lbrack y(\tau ),t(\tau )]=\phi
^{h}[y(\tau ),t(\tau )]-x(\tau ).$ Putting this in the right hand side of
the particle equation of motion (\ref{5.6}), we obtain a quantum Langevin
equation, 

\begin{equation}
m{\frac{d^{2}x}{d\tau ^{2}}}+\zeta {\frac{dx}{d\tau }}+Kx=F(\tau ),
\label{5.10}
\end{equation}

where $\zeta $ is the friction constant, given by the same expression (\ref
{4.6}) obtained for the oscillator at a fixed point. In this Langevin
equation, the fluctuating operator force $F(\tau )$ is given by 

\begin{equation}
F(\tau )=\zeta {\frac{d\phi ^{h}[y(\tau ),t(\tau )]}{d\tau }}.  \label{5.11}
\end{equation}

We note that this Langevin equation has the same form as the Langevin
equation (\ref{4.5}) corresponding to the oscillator at a fixed point.
Indeed, it reduces to that equation for the special motion $y(\tau )=0$, $
t(\tau )=t$.

Next consider the correlation of the fluctuating force. Using the above
definition, we see that 

\begin{equation}
{\frac{1}{2}}\left\langle F(\tau _{1})F(\tau _{2})+F(\tau _{2})F(\tau
_{1})\right\rangle =\zeta ^{2}{\frac{d^{2}}{d\tau _{1}d\tau _{2}}}C[y(\tau
_{1})-y(\tau _{2}),t(\tau _{1})-t(\tau _{2})],  \label{5.12}
\end{equation}

where $C(\Delta y,\Delta t)$ is the correlation function (\ref{2.13}) for
the real scalar field. At $T=0$, this correlation is given by the explicit
expression (\ref{2.25}), so we can write 

\begin{equation}
{\frac{1}{2}}\left\langle F(\tau _{1})F(\tau _{2})+F(\tau _{2})F(\tau
_{1})\right\rangle ={\frac{\hbar \zeta }{2\pi }}{\frac{d^{2}}{d\tau
_{1}d\tau _{2}}}\log (\Delta t^{2}-\Delta y^{2}/c^{2}).\quad (T=0)
\label{5.13}
\end{equation}

The commutator of the fluctuating force is given by 

\begin{equation}
\lbrack F(\tau _{1}),F(\tau _{2})]=\zeta ^{2}{\frac{d^{2}}{d\tau _{1}d\tau
_{2}}}[\phi (y_{1},t_{1}),\phi (y_{2},t_{2})]  \label{5.14}
\end{equation}

Using the explicit expression (\ref{2.29}), we can write 

\begin{equation}
\lbrack F(\tau _{1}),F(\tau _{2})]=-i\hbar \zeta {\frac{d^{2}}{d\tau
_{1}d\tau _{2}}}\text{sgn}(\Delta \tau )=2i\hbar \zeta \delta ^{\prime
}(\tau _{1}-\tau _{2}).  \label{5.15}
\end{equation}

Note that the form of the Langevin equation and the commutator of the
fluctuating force operator are independent of the motion. However, the
correlation of the force is explicitly dependent upon the motion.

Now we consider the special case of hyperbolic motion. We could use the
explicit expression (\ref{5.13}) for the force correlation, but it will be
useful in the later discussion to obtain the expression in a different form.
We begin with the normal mode expansion (\ref{4.8}) of the free field, which
with the expression (\ref{5.11}) for the fluctuating force results in the
expansion

\begin{equation}
F(\tau )={\frac{d}{d\tau }}\sum_{k}\sqrt{{\frac{\zeta \hbar c}{L\omega }}}
(a_{k}e^{i[ky(\tau )-\omega t(\tau )]}+a_{k}^{\dag }e^{-i[ky(\tau )-\omega
t(\tau )]}).  \label{5.16}
\end{equation}

We next introduce the Fourier expansion

\begin{equation}
e^{i[ky(\tau )-\omega t(\tau )]}={\frac{1}{2\pi }}\int_{-\infty }^{\infty
}d\omega ^{\prime }c(k;\omega ^{\prime })e^{-i\omega ^{\prime }\tau },
\label{5.17}
\end{equation}

to write

\begin{equation}
F(\tau )={\frac{1}{2\pi i}}\int_{-\infty }^{\infty }d\omega ^{\prime }\omega
^{\prime }\sum_{k}\sqrt{{\frac{\zeta \hbar c}{L\omega }}}(a_{k}c(k;\omega
^{\prime })e^{-i\omega ^{\prime }\tau }-a_{k}^{\dag }c(k;\omega ^{\prime
})^{\ast }e^{i\omega ^{\prime }\tau }).  \label{5.18}
\end{equation}

Forming the zero temperature correlation, using the expectation values (\ref%
{2.12}) we can write

\begin{eqnarray}
{\frac{1}{2}}\left\langle F(\tau _{1})F(\tau _{2})+F(\tau _{2})F(\tau
_{1})\right\rangle &=&{\frac{\zeta }{4\pi ^{2}}}\int_{-\infty }^{\infty
}d\omega _{1}\int_{-\infty }^{\infty }d\omega _{2}\omega _{1}\omega _{2} 
\nonumber \\
&&\mathrm{Re}\left\{ e^{-i(\omega _{1}\tau _{1}-\omega _{2}\tau
_{2})}\sum_{k}\frac{\hbar c}{L\omega }c(k;\omega _{1})c(k;\omega _{2})^{\ast
}\}\right\} .  \label{5.19}
\end{eqnarray}

To evaluate this expression, we first consider the Fourier transform,

\begin{equation}
c(k;\omega ^{\prime })=\int_{-\infty }^{\infty }d\tau e^{i[\omega ^{\prime
}\tau +ky(\tau )-\omega t(\tau )]}.  \label{5.20}
\end{equation}

Note first that for hyperbolic motion $y(\tau )$ is even and $t(\tau )$ is
odd as a function of $\tau $, so

\begin{equation}
c(-k;\omega ^{\prime })=c(k;\omega ^{\prime })^{\ast }.  \label{5.21}
\end{equation}

It is therefore sufficient to consider positive $k=\omega /c$. For this
case, using the equations (\ref{4.8}) of hyperbolic motion, we make in the
integral (\ref{5.17}) the substitution $z=e^{-F\tau /mc}$ to get

\begin{equation}
c(\frac{\omega }{c};\omega ^{\prime })=\frac{mc}{F}\int_{0}^{\infty
}dzz^{-1-imc\omega ^{\prime }/F}e^{-mc\omega z/F}.  \label{5.22}
\end{equation}

Finally, we rotate the path of integration into the positive imaginary axis,
and use the well known integral representation of the gamma function
\cite{bateman_it1}, to obtain the result

\begin{equation}
c(\frac{\omega }{c};\omega ^{\prime })={\frac{mc}{F}}\left( {\frac{mc\omega 
}{F}}\right) ^{imc\omega ^{\prime }/F}e^{\pi mc\omega ^{\prime }/2F}\Gamma
(-i\frac{mc\omega ^{\prime }}{F}).  \label{5.23}
\end{equation}

Next consider

\begin{eqnarray}
\sum_{k}{\frac{\hbar c}{L\omega }}c(k;\omega _{1})c(k;\omega _{2})^{\ast }
&=&\frac{\hbar m^{2}c^{2}}{\pi F^{2}}e^{\pi mc(\omega _{1}+\omega
_{2})/2F}\Gamma (-i\frac{mc\omega _{1}}{F})\Gamma (i\frac{mc\omega _{2}}{F})
\nonumber \\
&&\times \mathrm{Re}\{\int_{0}^{\infty }d\omega {\frac{1}{\omega }}\left( {
\frac{mc\omega }{F}}\right) ^{imc(\omega _{1}-\omega _{2})/F}\}  \label{5.24}
\end{eqnarray}

where we have used the prescription (\ref{2.15}) for the limit $L\rightarrow
\infty $, then the condition.(\ref{5.21}) and the dispersion relation (\ref
{2.10}) to to write the integral as over positive frequencies. With the
substitution $v=\log (mc\omega /F)$ the integral here becomes the familiar
integral for the Dirac delta-function and is therefore given by $(2\pi
F/mc)\delta (\omega _{1}-\omega _{2})$. It follows that 

\begin{equation}
\sum_{k}{\frac{\hbar c}{L\omega }}c(k;\omega _{1})c(k;\omega _{2})^{\ast
}=2\pi \hbar \frac{e^{\pi mc\omega _{1}/F}}{\omega _{1}\sinh \frac{\pi
mc\omega _{1}}{F}}\delta (\omega _{1}-\omega _{2}),  \label{5.25}
\end{equation}

where we have used the identity $\left\vert \Gamma (ix)\right\vert ^{2}=\pi
/x\sinh \pi x$. Using this result in Eq. (\ref{5.19}), we find that the
zero-temperature correlation can be expressed in the form 

\begin{equation}
{\frac{1}{2}}\left\langle F(\tau _{1})F(\tau _{2})+F(\tau _{2})F(\tau
_{1})\right\rangle ={\frac{\hbar }{\pi }}\int_{0}^{\infty }d\omega \zeta
\omega \coth \frac{\pi mc\omega }{F}\cos [\omega (\tau _{1}-\tau _{2})].
\label{5.26}
\end{equation}

Again, we see the Unruh temperature (\ref{3.6}). That is, this force
autocorrelation seen by the oscillator in hyperbolic motion through a zero
temperature field is identical with that (\ref{4.10}) seen by an oscillator
at rest in a field at the Unruh temperature. (Recall that the proper time $
\tau $ is the time as measured on a clock moving with the oscillator)We
emphasize that this means that the moving oscillator is itself at the Unruh
temperature.

\section{Energy radiated by an oscillator undergoing hyperbolic motion.}

\label{sec:six}

We now calculate the flux of energy radiated by the oscillator undergoing
hyperbolic motion in a zero-temperature field. As we have seen, the moving
oscillator is at the elevated Unruh temperature. This picture of a hot
oscillator moving through a zero-temperature background leads one to expect
that there should be radiation. After all, doesn't heat always flow from a
hot body to a cold body? But in this section we shall show by explicit
calculation that the net energy flux is zero.

Consider now the flux of field energy radiated by the oscillator as measured
at some point to the left of the point of closest approach in hyperbolic
motion. The energy flux operator is given by (\ref{2.6}) with $
u(y,t)\rightarrow \phi (y,t)$, the total field given by the solution (\ref
{5.8}) of the field equations for the oscillator in hyperbolic motion.
Forming the expectation, we can write just as in Sec. \ref{sec:four},

\begin{equation}
\left\langle j(y,t)\right\rangle =\left\langle j_{0}(y,t)\right\rangle
+\left\langle j_{\mathrm{dir}}(y,t)\right\rangle +\left\langle j_{\mathrm{int
}}(y,t)\right\rangle ,  \label{6.1}
\end{equation}

where $\left\langle j_{0}(y,t)\right\rangle $ is the energy flux (\ref{4.15}
) in the absence of the oscillator, while now the direct flux is given by

\begin{equation}
\left\langle j_{\mathrm{dir}}(y,t)\right\rangle =-\frac{\zeta c}{2}\mathrm{Re
}\left\langle \frac{\partial x(\tau ^{\mathrm{ret}})}{\partial t}\frac{
\partial x(\tau ^{\mathrm{ret}})}{\partial y}\right\rangle ,  \label{6.2}
\end{equation}

and the interference term is given by

\begin{equation}
\left\langle j_{\mathrm{int}}(y,t)\right\rangle =\frac{\zeta c}{2}\mathrm{Re}
\left\langle \frac{\partial \phi ^{h}(y,t)}{\partial t}\frac{\partial x(\tau
^{\mathrm{ret}})}{\partial y}+\frac{\partial x(\tau ^{\mathrm{ret}})}{
\partial t}\frac{\partial \phi ^{h}(y,t)}{\partial y}\right\rangle .
\label{6.3}
\end{equation}

In these expressions, the retarded time is determined by the condition (\ref
{5.9}). For a field point to the left of the point of closest approach, $
y<mc^{2}/F$ and, using the equations (\ref{3.2}) of hyperbolic motion, we
find

\begin{equation}
\tau ^{\mathrm{ret}}=\frac{mc}{F}\log \frac{F(t+y/c)}{mc}.  \label{6.4}
\end{equation}

Thus $c\partial \tau ^{\mathrm{ret}}/\partial y=\partial \tau ^{\mathrm{ret}
}/\partial t$ and using the chain rule we can write

\begin{equation}
\left\langle j_{\mathrm{dir}}(y,t)\right\rangle =-\frac{1}{2}\left( \frac{
\partial \tau ^{\mathrm{ret}}}{\partial t}\right) ^{2}\zeta \left\langle
\left( \frac{dx(\tau ^{\mathrm{ret}})}{d\tau ^{\mathrm{ret}}}\right)
^{2}\right\rangle  \label{6.5}
\end{equation}

and

\begin{equation}
\left\langle j_{\mathrm{int}}(y,t)\right\rangle =\frac{\zeta }{2}\frac{
\partial \tau ^{\mathrm{ret}}}{\partial t}\mathrm{Re}\left\langle \frac{
dx(\tau ^{\mathrm{ret}})}{d\tau ^{\mathrm{ret}}}\left( \frac{\partial \phi
^{h}(y,t)}{\partial t}+c\frac{\partial \phi ^{h}(y,t)}{\partial y}\right)
\right\rangle .  \label{6.6}
\end{equation}

As in Sec. \ref{sec:four}, we get some insight into the significance of
these terms if we consider the energy balance equation for the moving
oscillator,

\begin{equation}
\frac{d}{d\tau }\left\langle \frac{1}{2}\left( {\frac{dx(\tau )}{d\tau }}
\right) ^{2}+\frac{1}{2}Kx^{2}(\tau )\right\rangle +\zeta \left\langle
\left( {\frac{dx(\tau )}{d\tau }}\right) ^{2}\right\rangle ={\frac{1}{2}}
\left\langle {\frac{dx(\tau )}{d\tau }}F(\tau )+F(\tau ){\frac{dx(\tau )}{
d\tau }}\right\rangle .  \label{6.7}
\end{equation}

This is identical with the corresponding energy balance equation for the
oscillator at rest, the only difference being that here $\tau $ is the time
as measured on a clock moving with the oscillator. Thus, in a frame moving
with the oscillator, the energy balance is identical with that for an
oscillator at rest. In particular, we can interpret the second term on the
left as the mean rate of radiation of field energy by the oscillator, while
the right hand side is the rate at which energy is supplied to the
oscillator by the fluctuating force, all as seen in the moving frame. The
direct flux (\ref{6.5}) is half the rate of radiation of energy multiplied
by the time dilation factor $(\partial \tau ^{\mathrm{ret}}/\partial t)^{2}$
. The factor of two is accounted for by the fact that half the radiation is
to the left, half to the right. The time dilation factor corresponds to the
transformation from the proper time kept on a clock moving with the
oscillator, where the rate of radiation is uniform, to a stationary clock at
the field point. Using the expression (\ref{6.4}) for the retarded time, we
see that

\begin{equation}
{\frac{\partial \tau ^{\mathrm{ret}}}{\partial t}}={\frac{mc}{F(t+y/c)}}
,\quad 0<t+y/c<\infty .  \label{6.8}
\end{equation}

Thus, $\left\langle j_{\mathrm{dir}}(y,t)\right\rangle$ corresponds to a
flux that is zero for $t<-y/c$, then suddenly infinite and decaying to zero
for long times. Finally, since $\left\langle j_{\mathrm{dir}
}(y,t)\right\rangle $ corresponds to the energy lost by the oscillator
through radiation, the interference term $\left\langle j_{\mathrm{int}
}(y,t)\right\rangle $ must be the inward flux of field energy absorbed by
the oscillator. Since the oscillator is in a stationary equilibrium state
corresponding to the Unruh temperature, one should expect that these two
fluxes should balance, just as they do for an oscillator at rest. We next
show by explicit calculation that these two fluxes do indeed cancel to give
a net flux of zero.

We begin using the expression (\ref{5.18}) for the fluctuating force, to
write the solution of the quantum Langevin equation (\ref{5.10}) in the form

\begin{equation}
x(\tau )={\frac{1}{2\pi i}}\int_{-\infty }^{\infty }d\omega ^{\prime }\omega
^{\prime }\sum_{k}\sqrt{{\frac{\zeta \hbar c}{L\omega }}}(a_{k}\alpha
(\omega ^{\prime })c(k;\omega ^{\prime })e^{-i\omega ^{\prime }\tau
}-a_{k}^{\dag }\alpha (\omega ^{\prime })^{\ast }c(k;\omega ^{\prime
})^{\ast }e^{i\omega ^{\prime }\tau }),  \label{6.9}
\end{equation}

where $\alpha (\omega )$ is the oscillator susceptibility (\ref{4.12}). With
this, forming the expectation using the expectation values (\ref{2.12}) with 
$T=0$, then using the result (\ref{5.25}) we find

\begin{equation}
\zeta \left\langle \left( \frac{dx(\tau )}{d\tau }\right) ^{2}\right\rangle =
{\frac{\hbar }{\pi }}\int_{0}^{\infty }d\omega \omega ^{3}\zeta
^{2}\left\vert \alpha (\omega )\right\vert ^{2}\coth \frac{\pi mc\omega }{F}.
\label{6.10}
\end{equation}

As we have seen, this is the rate at which the oscillator loses energy
through radiation It is independent of time as measured in the moving frame
and identical with the same quantity for a stationary oscillator. With this,
we see that the direct flux can be written

\begin{equation}
\left\langle j_{\mathrm{dir}}(y,t)\right\rangle =-\left( \frac{\partial \tau
^{\mathrm{ret}}}{\partial t}\right) ^{2}{\frac{\hbar }{2\pi }}
\int_{0}^{\infty }d\omega \omega ^{3}\zeta ^{2}\left\vert \alpha (\omega
)\right\vert ^{2}\coth \frac{\pi mc\omega }{F}.  \label{6.11}
\end{equation}

Next consider the interference term (\ref{6.6}). Using the expansion (\ref
{4.8}) of the free field and the expression (\ref{6.9}) for the oscillator
displacement, we form the expectation using the expectation values (\ref
{2.12}) with $T=0$ to get

\begin{eqnarray}
\left\langle j_{\mathrm{int}}(y,t)\right\rangle &=&\frac{\zeta }{2}\frac{
\partial \tau ^{\mathrm{ret}}}{\partial t}\mathrm{Im}{\frac{1}{2\pi }}
\int_{-\infty }^{\infty }d\omega ^{\prime }\omega ^{\prime 2}\alpha (\omega
^{\prime })e^{-i\omega ^{\prime }\tau ^{\mathrm{ret}}}  \nonumber \\
&&\times \frac{\hbar c}{L}\sum_{k}\frac{\omega -ck}{\omega }c(k;\omega
^{\prime })e^{-i(ky-\omega t)}.  \label{6.12}
\end{eqnarray}

Since $\omega =c\left\vert k\right\vert $, the sum can be restricted to
negative $k$. Then, replacing the sum by an integral, using the prescription
(\ref{2.15}) and using the identity (\ref{5.21}) and the expression (\ref
{5.23}) for $c(\frac{\omega }{c};\omega ^{\prime })$ we get

\begin{equation}
\frac{\hbar c}{L}\sum_{k}\frac{\omega -ck}{\omega }c(k;\omega ^{\prime
})e^{-i(ky-\omega t)}=e^{\pi mc\omega ^{\prime }/2F}\Gamma (i\frac{mc\omega
^{\prime }}{F})\frac{mc\hbar }{F\pi }\int_{0}^{\infty }d\omega \left( {\frac{
mc\omega }{F}}\right) ^{-imc\omega ^{\prime }/F}e^{i\omega (t+y/c)}.
\label{6.13}
\end{equation}

Next, we rotate the path of integration into the positive real axis and use
the integral representation of the gamma function \cite{bateman_it1} to
write

\begin{eqnarray}
\frac{\hbar c}{L}\sum_{k}\frac{\omega -ck}{\omega }c(k;\omega ^{\prime
})e^{-i(ky-\omega t)} &=&\frac{mc}{F(t+y/c)}({\frac{mc}{F(t+y/c)})}
^{-imc\omega ^{\prime }/F}  \nonumber \\
&&\times e^{\pi mc\omega ^{\prime }/F}\frac{i\hbar }{\pi }\Gamma (i\frac{
mc\omega ^{\prime }}{F})\Gamma (1-i\frac{mc\omega ^{\prime }}{F}).
\label{6.14}
\end{eqnarray}

We use in the first factor the expression (\ref{6.8}) for $d\tau ^{\mathrm{
ret}}/dt$ and in the second factor the expression (\ref{6.4}) for $\tau ^{
\mathrm{ret}}$. Then, using the identity $i\Gamma (iz)\Gamma (1-ix)=\pi
/\sinh \pi x$, we obtain the result

\begin{equation}
\frac{\hbar c}{L}\sum_{k}\frac{\omega -ck}{\omega }c(k;\omega ^{\prime
})e^{-i(ky-\omega t)}=\hbar {\frac{\partial \tau ^{\mathrm{ret}}}{\partial t}
}\frac{e^{\pi mc\omega ^{\prime }/F}}{\sinh \frac{mc\omega ^{\prime }}{F}}
e^{i\omega ^{\prime }\tau ^{\mathrm{ret}}}.  \label{6.15}
\end{equation}

Putting this in the expression (\ref{6.12}) we get

\begin{equation}
\left\langle j_{\mathrm{int}}(y,t)\right\rangle =\left( \frac{\partial \tau
^{\mathrm{ret}}}{\partial t}\right) ^{2}{\frac{\hbar \zeta }{2\pi }}
\int_{-\infty }^{\infty }d\omega \omega ^{2}\mathrm{Im}\{\alpha (\omega
)\}\coth \frac{\pi mc\omega }{F}.  \label{6.16}
\end{equation}

But, as we see from the expression (\ref{4.12}) for the oscillator
susceptibility, $\mathrm{Im\{}\alpha (\omega )\}=\omega \zeta \left\vert
\alpha (\omega )\right\vert ^{2}$. Therefore, this is just the negative of
the expression (\ref{6.11}) for $\left\langle j_{\mathrm{dir}
}(y,t)\right\rangle $. That is, 

\begin{equation}
\left\langle j_{\mathrm{dir}}(y,t)\right\rangle +\left\langle j_{\mathrm{int}
}(y,t)\right\rangle =0.  \label{6.17}
\end{equation}

Thus, the expected energy flux vanishes,

\begin{equation}
\left\langle j(y,t)\right\rangle =0.  \label{6.18}
\end{equation}

We conclude that a system that undergoes uniform acceleration with respect
to the vacuum of flat space-time does not radiate despite the fact that it
does in fact thermalize at the Unruh temperature.

\section{Concluding remarks}

\label{sec:seven}

A system undergoing hyperbolic motion through a zero-temperature vacuum
experiences a finite temperature, the Unruh temperature (\ref{3.6}). This
was pointed out by Davies \cite{davies75} and Unruh \cite{unruh76}. Our
explicit calculation for the scalar electrodynamics model verifies this for
an oscillator in hyperbolic motion. The effect is real: the moving
oscillator is in an equilibrium state identical with that of one at rest at
the Unruh temperature, with a corresponding distribution over excited states.

This picture of a system at a finite temperature moving through a
zero-temperature vacuum might lead one to expect that there would be energy
radiated. However Grove \cite{grove86} and Raine et al.\cite{raine91}
argued that this was not the case, there is no radiation. In agreement with
them, our explicit model calculation shows that there is in fact no
radiation of energy. The situation is exactly the same as that for a system
at rest in a zero-temperature vacuum. The system is driven by the zero-point
oscillations of the vacuum field while simultaneously radiating energy into
the vacuum. But the driving force and the radiation reaction exactly
balance, so the system remains in equilibrium with no net radiation of
energy. For our simple model of an oscillator with scalar electrodynamics we
show by explicit calculation that the net flux of radiant energy at a point
in the field away from the oscillator is zero, for an oscillator at rest in
Sec. \ref{sec:four} and for an oscillator in hyperbolic motion in Sec. \ref
{sec:six}. The fact that the argument is identical for an oscillator at
rest and one in hyperbolic motion makes it difficult to escape the
conclusion that, on very general grounds, there is no radiation in either case.
You can't have the one without the other.

We have seen in Sec. \ref{sec:four} that when the oscillator is excited by
an external agent it will radiate, since the externally excited motion is
uncorrelated with the fluctuations of the vacuum field. \ Unruh \cite
{unruh92}, in his response to the paper of Raine et al.\cite{raine91},
introduces a heat bath moving with the oscillator. This bath is assumed to
be at the Unruh temperature and when it drives the oscillator there will be
radiation, the bath acting as an external agent. We have serious
reservations about this picture, but whatever its merits, it certainly does
not represent the situation envisioned in the many proposals to observe the
radiation, all of which involve a single particle or at most a single atomic
system in accelerated motion.  Moreover, Unruh places emphasis on " - - the
radiation  - - expected from the oscillator/heat bath coming into equilibrium
with the thermal radiation in the far past" \cite{unruh92}.  Next, Parentani
\cite{par1995} expands on this discussion and shows explicitly by an " - -
analysis of the transients when one switches off the interaction - -." that such
transients lead to radiation. However, during the switching-on and switching-off
of the external force, we have a situation which is outside the realm of what is
understood to be the basis for Unruh radiation.  

Of course, hyperbolic motion is an idealization, with the force $F$ applied
over an infinite time. More realistically one could assume that at some
distant but finite time in the past the oscillator is impulsively
accelerated into hyperbolic motion and the constant force is switched on. At
that time there must be an exchange of energy with the field, but it would
not be what one would call Unruh radiation. A description of this exchange
is outside the range of the present discussion.

Our conclusion is that a system in hyperbolic motion through a
zero-temperature vacuum does not radiate, despite the fact that it is in a
state corresponding to the elevated Unruh temperature. Our conclusion is that a
system in hyperbolic motion through a zero-temperature vacuum does not radiate,
despite the fact that it is in a state corresponding to the elevated Unruh
temperature.  We should point out that it has been argued by some authors
\cite{van2001}, \cite{hig1992} that this is an artifact of the model we have
used.  In particular, the interaction of a charged oscillator with the
electromagnetic field was discussed by Vanzella et al. \cite{van2001} and the
authors conclude that there is radiation. However, we are skeptical since, as we
have remarked above, the argument is essentially one of detailed balance: for a
system in equilibrium the rate of emission of radiation is exactly balanced by a
corresponding absorption, there is no net radiation.  What we have done here is
to demonstrate in detail that detailed balance holds for a system in hyperbolic
motion exactly as it does for a system at rest at a finite temperature.  It is
difficult to believe that this principle is model-dependent.

\end{document}